\documentclass[aps,final,tightenlines,onecolumn,11pt,notitlepage]{revtex4-1}
\usepackage{amsmath}
\usepackage{amssymb}
\usepackage[mathcal]{euscript}
\usepackage{xfrac}
\usepackage{bm}
\usepackage{lineno}

\usepackage{graphicx}   % need for figures
\graphicspath{{./pics/}}

\usepackage{color}      % use if color is used in text
\usepackage{subfigure}  % use for side-by-side figures

\usepackage{verbatim}   % useful for program listings
\usepackage{algorithm}
\usepackage[noend]{algpseudocode}

\begin{document}

\title{Parameter Estimation for Subgrid-Scale Models Using Markov Chain Monte Carlo Approximate Bayesian Computation}
\author{Olga A.~Doronina}
\affiliation{Paul M.~Rady Department of Mechanical Engineering, University of Colorado, Boulder, CO 80309, USA}
\author{Colin A.~Z.~Towery}
\affiliation{Paul M.~Rady Department of Mechanical Engineering, University of Colorado, Boulder, CO 80309, USA}
\author{Peter E.~Hamlington}
\affiliation{Paul M.~Rady Department of Mechanical Engineering, University of Colorado, Boulder, CO 80309, USA}
\email{peh@colorado.edu}

%%%%%%%%%%%%%%%%%%%%%%%%%%%%%%%%
% Abstract
%%%%%%%%%%%%%%%%%%%%%%%%%%%%%%%%
\begin{abstract}
We use approximate Bayesian computation (ABC) combined with an ``improved'' Markov chain Monte Carlo (IMCMC) method to estimate posterior distributions of model parameters in subgrid-scale (SGS) closures for large eddy simulations (LES) of turbulent flows. The ABC-IMCMC approach avoids the need to directly compute a likelihood function during the parameter estimation, enabling a substantial speed-up and greater flexibility as compared to full Bayesian approaches. The method also naturally provides uncertainties in parameter estimates, avoiding the artificial certainty implied by many optimization methods for determining model parameters. In this study, we outline details of the present ABC-IMCMC approach, including the use of an adaptive proposal and a calibration step to accelerate the parameter estimation process. We demonstrate the approach by estimating parameters in two nonlinear SGS closures using reference data from direct numerical simulations of homogeneous isotropic turbulence. We show that the resulting parameter values give excellent agreement with reference probability density functions of the SGS stress and kinetic energy production rate in \emph{a priori} tests, while also providing stable solutions in forward LES (i.e., \emph{a posteriori} tests) for homogeneous isotropic turbulence. The ABC-IMCMC method is thus shown to be an effective and efficient approach for estimating unknown parameters, including their uncertainties, in SGS closure models for LES of turbulent flows.
\end{abstract}

%%%%%%%%%%%%%%%%%%%%%%%%%%%%%%%%
% Keywords
%%%%%%%%%%%%%%%%%%%%%%%%%%%%%%%%

\maketitle
%\linenumbers

%%%%%%%%%%%%%%%%%%%%%%%%%%%%%%%%
% Sec: Introduction
%%%%%%%%%%%%%%%%%%%%%%%%%%%%%%%%

\section{Introduction}
%Setup the problem
Large eddy simulations (LES) have the potential to provide a nearly ideal blend of physical accuracy and computational cost for the simulation of turbulent flows, but the predictive power of such simulations depends on the accuracy of closure models for unresolved subgrid scale (SGS) fluxes. Attempts to develop such models from physics principles alone have thus far failed to yield a universally accurate SGS model. Consequently, the overwhelming majority of LES continues to be performed using classical \cite{smagorinsky_general_1963} or dynamic \cite{germano_dynamic_1991} Smagorinsky models, or with artificial kinetic energy dissipation produced by low-order numerical schemes (i.e., implicit LES). Although such simple or purely numerical models are robust and provide stable LES solutions, they typically perform poorly in flows with complex physics (e.g., combustion) and in situations where the fundamental principles underlying LES break down (e.g., in regions near solid boundaries where there is no longer a clear separation between kinetic energy input and dissipation scales). At the same time, attempts to develop more sophisticated models are typically plagued by the presence of many unknown model parameters, which can be difficult to simultaneously calibrate across different flows.
	
%The issue with optimization/inversion approaches
Traditionally, model parameters have been determined using either optimization techniques or direct inversion of model equations given some reference (e.g., experimental or higher-fidelity simulation) data. For example, successful attempts have been made to find the optimal Smagorinsky constant using an optimization technique~\cite{meyers2006optimal}, Kriging-based response surfaces~\cite{jouhaud2008sensitivity}, and neural networks~\cite{sarghini_neural_2003}, and direct inversion techniques have been used to estimate model parameters for both Reynolds averaged Navier Stokes (RANS) simulations \cite{parish_paradigm_2016} and for LES \cite{Wang2005}. However, Oberkampf, Trucano, and co-authors \cite{oberkampf2000validation,oberkampf2002verification,oberkampf2004verification,oberkampf2010verification} have advocated caution when using optimization approaches, noting that model parameter estimates should also include a quantification of uncertainty. This is especially true given the uncertain nature of essentially all reference data (even if only due to statistical non-convergence), as well as the approximate nature of SGS models (even for the most sophisticated models). Direct inversion approaches can also become challenging (although not impossible) for complex model forms or when the model itself consists of partial differential equations. 

%The issue with full Bayesian approaches
Statistical methods, such as Bayesian approaches, provide an alternative path to model parameter calibration, giving a posterior probability distribution of unknown parameters. For example, \citet{oliver_bayesian_2011}, \citet{cheung_bayesian_2011}, \citet{ray_bayesian_2014, ray_bayesian_2016}, \citet{xiao2016, xiao_review_2019}, and \citet{zhang_efficient_2019} have all applied Bayesian methods to Reynolds stress models for RANS simulations in order to estimate model parameters and uncertainties. Recently, \citet{safta_uncertainty_2017} used a Bayesian approach to estimate a joint distribution for LES SGS model parameters. A benefit of the Bayesian statistical approach is that the posterior probability density also naturally provides uncertainties associated with each estimated parameter, in contrast to other inversion techniques that provide only single value estimates for unknown parameters. However, solving the full Bayesian problem requires knowledge of the likelihood function, which can be either difficult or costly to compute, or both. In many cases, this likelihood function is approximated using a Gaussian formulation, which is not valid in general for all model forms and all flows. 

%Introduce ABC
In this paper, we outline the use of approximate Bayesian computation (ABC) and an ``improved'' Markov chain Monte Carlo (IMCMC) method to determine unknown model parameters and their uncertainties. Most significantly, the ABC method approximates the posterior distribution of parameters without using a likelihood function. The ABC method was introduced and first widely applied in population genetics~\cite{beaumont2002,marjoram_markov_2003,wegmann_efficient_2009} and molecular genetics~\cite{marjoram2006}. It has subsequently been used in other scientific areas such as astrophysics~\cite{wawrzynczak2018,cameron2012}, chemistry~\cite{picchini2014}, epidemiology~\cite{luciani2009epidemiological,zheng2017} and ecology~\cite{beaumont2010}. More detailed recent reviews of the ABC approach are provided by~\citet{csillery2010}, \citet{marin_approximate_2012}, \citet{lintusaari2017fundamentals}, \citet{sisson2018handbook} and, most recently, \citet{beaumont2019}. The ABC method has also previously been used in engineering contexts for the estimation of kinetic rate coefficients for chemical reaction mechanisms \cite{khalil2018} and for the estimation of boundary conditions in complex thermo-fluid flows \cite{christopher2017aiaa_scitech,christopher2017aiaa_aviation,christopher2018}. Here, we are the first to take advantage of both ABC and IMCMC for discovering model parameter values and uncertainties in multi-parameter SGS closures. 

%Outline the paper
In the following, the inverse problem solved by the ABC-IMCMC approach is outlined in Section \ref{sec:inverse} and details of the ABC-IMCMC approach, including the use of an adaptive proposal and a calibration step, are described in Sections \ref{sec:method} and \ref{sec:details}. Results are presented in Section \ref{sec:results}, and conclusions and directions for future work are provided at the end.

%%%%%%%%%%%%%%%%%%%%%%%%%%%%%%%%
% Sec: Inverse Problem
%%%%%%%%%%%%%%%%%%%%%%%%%%%%%%%%
\section{Subgrid-Scale Modeling as an Inverse Problem\label{sec:inverse}}
Coarse-graining of the Navier-Stokes equations using a low-pass filter, denoted $\widetilde{(\cdot)}$, at scale $\Delta$ gives the LES equations for an incompressible flow, which are written as \cite{meneveau2000}
\begin{align}
    \frac{\partial \widetilde{u}_i}{\partial x_i} &= 0\,,\\
	\frac{\partial \widetilde{u}_i}{\partial t} + \widetilde{u}_j \frac{\partial \widetilde{u}_i}{\partial x_j} &= -  \frac{\partial \widetilde{p}}{\partial x_i} + \nu \frac{\partial^2 \widetilde{u}_i}{\partial x_j \partial x_j} - \frac{\partial \tau_{ij}}{\partial x_j}\,,
	\label{eq:nse}
\end{align}
where $\widetilde{u}_i$ is the resolved-scale velocity, $\widetilde{p}$ is the resolved-scale pressure normalized by density, $\nu$ is the kinematic viscosity, and $\tau_{ij}$ is the unclosed SGS stress tensor given by
\begin{equation}
	\label{eq:tau}
	\tau_{ij} = \widetilde{u_i u_j} - \widetilde{u}_i \widetilde{u}_j\,.
\end{equation}
The LES scale $\Delta$ is often termed the grid scale since this is the finest scale represented when using the grid discretization as an implicit LES filter. More broadly, $\Delta$ is the filter cut-off scale, whether that filter is applied explicitly or implicitly. The SGS stress $\tau_{ij}$ prevents closure and, in order to solve Eq.~\eqref{eq:nse}, an appropriate relation for $\tau_{ij}$ must be found in terms of resolved-scale quantities only.
	
Closure of Eq.~\eqref{eq:nse} can be achieved by modeling the deviatoric part of the stress tensor $\sigma_{ij} = \tau_{ij} - \tau_{kk}(\delta_{ij}/3)$, which, it is assumed, can be approximated by an unknown, high dimensional, non-parametric functional $\mathcal{F}_{ij}$ that takes as its arguments only quantities that can be expressed in terms of the resolved-scale strain rate, $\widetilde{S}_{ij}$, and rotation rate, $\widetilde{R}_{ij}$, tensors \cite{pope_more_1975}; namely 
\begin{align}
	\sigma_{ij}(\textbf{x},t)\approx \mathcal{F}_{ij}\left[ \widetilde{S}_{ij}(\textbf{x}+\textbf{x}',t-t'),\widetilde{R}_{ij}(\textbf{x}+\textbf{x}',t-t') \right]\,,
	\label{eq:fund}
\end{align}
for all $\textbf{x}'$ and $t'\geq 0$, where $\widetilde{S}_{ij}$ and $\widetilde{R}_{ij}$ are given by
\begin{equation}\label{eq:SR_LES}
	\widetilde{S}_{ij} = \frac{1}{2} \left( \frac{\partial \widetilde{u}_i }{\partial x_j} + \frac{\partial \widetilde{u}_j}{ \partial x_i} \right)\,,\quad \widetilde{R}_{ij} = \frac{1}{2} \left( \frac{\partial \widetilde{u}_i }{\partial x_j} - \frac{\partial \widetilde{u}_j}{ \partial x_i} \right)\,.
\end{equation}
It should be noted that the closure relation in Eq.~\eqref{eq:fund} allows SGS stresses at location $\textbf{x}$ and time $t$ to depend on $\widetilde{S}_{ij}$ and $\widetilde{R}_{ij}$, as well as their products, at any point in the flow and at any prior time. 

As a demonstration of the ABC approach for determining SGS model parameters, here we will use a single-point, single-time nonlinear model introduced by \citet{pope_more_1975}, which is given as
\begin{equation}\label{eq:closure}
	\mathcal{F}_{ij}(\bm{c}) = \sum_{n=1}^{4} c_{n} \widetilde{G}_{ij}^{(n)}\,,
\end{equation}
where $c_{n}$ are non-dimensional coefficients (i.e., the unknown model ``parameters'') that can depend on invariants of $\widetilde{S}_{ij}$ and $\widetilde{R}_{ij}$, and $\widetilde{G}_{ij}^{(n)}$ are tensor bases formed from products of $\widetilde{S}_{ij}$ and $\widetilde{R}_{ij}$ up to second order,
\begin{align}
	& \widetilde{G}_{ij}^{(1)} = \Delta^2 |\widetilde{S}| \widetilde{S}_{ij}\,, \\
	& \widetilde{G}_{ij}^{(2)} = \Delta^2\left(\widetilde{S}_{ik}\widetilde{R}_{kj} - \widetilde{R}_{ik}\widetilde{S}_{kj}\right)\,, \\
	& \widetilde{G}_{ij}^{(3)} = \Delta^2\left(\widetilde{S}_{ik}\widetilde{S}_{kj} - \frac{1}{3}\delta_{ij} \widetilde{S}_{kl}\widetilde{S}_{kl}\right) \,, \\
	& \widetilde{G}_{ij}^{(4)} = \Delta^2\left(\widetilde{R}_{ik}\widetilde{R}_{kj} - \frac{1}{3}\delta_{ij} \widetilde{R}_{kl}\widetilde{R}_{kl}\right)\,,
\end{align}
where $|\widetilde{S}| \equiv \left( 2\widetilde{S}_{ij} \widetilde{S}_{ij}\right)^{1/2}$. It should be noted that the original formulation by~\citet{pope_more_1975} extends to fifth order, but here the model is truncated for simplicity in order to demonstrate the ABC-IMCMC method. The first tensor basis corresponds to the classical Smagorinsky model, whereas the remaining bases represent second-order nonlinear corrections to this classical model. As outlined by \citet{king2016}, the non-parametric functional $\mathcal{F}_{ij}$ in Eq.~\eqref{eq:fund} can also be written as a Volterra series~\cite{boyd1985} or any other appropriate high-dimensional non-parametric functional. The exact non-parametric representation of $\mathcal{F}_{ij}$ can thus vary and is not fundamental to the present demonstration of the ABC-IMCMC approach for model parameter estimation.
	
Using data from experiments or direct numerical simulations (DNS), $\sigma_{ij}(\mathbf{x}, t)$ can be calculated and denoted as reference data $\mathcal{D}_{ij}$. Thus, we obtain the inverse problem
\begin{equation}
	\mathcal{F}_{ij}(\bm{c}) = \mathcal{D}_{ij}\,,
\end{equation}
where the model parameters of $\mathcal{F}_{ij}$ (i.e., $\bm{c}$) must be determined through an appropriate inversion technique given the data $\mathcal{D}_{ij}$. Using direct inversion, \citet{Wang2005} have demonstrated a nonlinear dynamic SGS model similar to that used here. Alternatively, the production rate field $\mathcal{P}=\sigma_{ij} \widetilde{S}_{ij}$ can instead be used as the reference data $\mathcal{D}_\mathcal{P}$, giving the inverse problem
\begin{equation}
	\mathcal{F}_{ij}(\bm{c})\widetilde{S}_{ij} = \mathcal{D}_\mathcal{P}\,.
\end{equation}
Although optimization techniques (e.g., as in \citet{king2016}) can be used to solve these inverse problems, the inversion process can be memory intensive and no uncertainty measures for the values of the unknown parameters in $\bm{c}$ are naturally provided. Moreover, the formulation of the inverse problem can become difficult for complex forms of $\mathcal{F}_{ij}$, including those involving non-algebraic relations (e.g., for models incorporating transport equations of turbulence quantities). 

In the following, we will demonstrate the use of the ABC-IMCMC approach to solve the inverse problems above, and we will compare estimates of the parameters $\bm{c}$ using reference data based on the stresses $\sigma_{ij}$ alone, the production $\mathcal{P}$ alone, and a combination of the stresses and production.

%%%%%%%%%%%%%%%%%%%%%%%%%%%%%%%%
% Sec: ABC Method
%%%%%%%%%%%%%%%%%%%%%%%%%%%%%%%%
\section{The ABC-IMCMC Parameter Estimation Method\label{sec:method}}
%========================================
% Subsec: Explanation of the ABC Approach
%========================================
\subsection{Approximate Bayesian computation}
Statistical inference is a powerful tool for solving inverse problems and Bayes' theorem, in particular, provides \textit{a posteriori} probability densities of model parameters $\bm{c}\in\bm{C}$ given reference data $\mathcal{D}$ as
\begin{equation}
	\label{eq:bayes_theorem}
	P(\bm{c}|\mathcal{D}) = \dfrac{L(\mathcal{D}\,|\,  \bm{c})\pi(\bm{c})}{\int_{\bm{C}}L(\mathcal{D}\,|\,  \bm{c})\pi(\bm{c})d\bm{c}}\,,
\end{equation}
where $L(\mathcal{D}\,|\, \bm{c})$ is the likelihood function and $\pi(\bm{c})$ is the prior distribution of model parameters. The posterior, $P(\bm{c}\,|\,\mathcal{D})$, provides the probability distribution of model parameters that satisfy the inverse problems outlined in the previous section, for given reference data. A benefit of the Bayesian statistical approach is that $P(\bm{c}\,|\,\mathcal{D})$ also naturally provides uncertainties associated with each estimated parameter, in contrast to other inversion techniques that provide only single-point estimates for unknown parameters.
	
The practical application of Bayes' theorem is often complicated by the fact that explicit analytical expressions for the likelihood function $L(\mathcal{D}\,|\, \bm{c})$ are rarely available. However, when the model parameter space $\bm{C}$ is finite and of low dimension, we can obtain the posterior density without an explicit likelihood function and without approximation using Algorithm~\ref{alg:rejection_sampling}, which was introduced by \citet{rubin_bayesianly_1984}. This algorithm samples model parameters $\bm{c}$ from a prior distribution $\pi(\bm{c})$ and compares model outcomes (i.e., model data) $\mathcal{D}^{\prime}$ with reference data $\mathcal{D}$, which may come from experiments or a higher fidelity model. The algorithm accepts parameters only if the modeled and reference data are exactly the same, and repeats until a total of $N$ model parameters have been accepted, from which the posterior joint probability density function (pdf) is then computed.

%algorithm-------------------------------
\begin{algorithm}[H]
	\caption{Bayesian rejection sampling algorithm~\cite{rubin_bayesianly_1984}}\label{alg:rejection_sampling}
	\begin{algorithmic}[1]
		\For{$i =1$ to $N$}
		\Repeat 
		\State Sample $\bm{c}_i$ from prior distribution $\pi(\bm{c})$
		\State Calculate $\mathcal{D}^{\prime} = \mathcal{F}(\bm{c}_i)$ from model
		\Until{$\mathcal{D}^{\prime} = \mathcal{D}$} 
		\State Accept $\bm{c}_i$
		\EndFor
		\State Using all accepted $\bm{c}_i$ calculate posterior joint pdf
	\end{algorithmic}
\end{algorithm}
%algorithm-------------------------------

For the determination of SGS model parameters, however, the parameter space $\bm{C}$ is continuous and the model is imperfect, making Algorithm~\ref{alg:rejection_sampling} impossible to use in the form outlined above since it is highly unlikely that $\mathcal{D}^{\prime}$ and $\mathcal{D}$ will ever be exactly the same. As an alternative, the acceptance criterion in Algorithm~\ref{alg:rejection_sampling} can be relaxed to
\begin{equation}
\label{key}
	d(\mathcal{D}^{\prime}, \mathcal{D}) \leq \varepsilon\,, 
\end{equation}
where $d$ is a distance function measuring the discrepancy between $\mathcal{D}^{\prime}$ and $\mathcal{D}$. That is, a sampled parameter value $\bm{c}$ is accepted as part of the posterior distribution if the corresponding distance $d$ is within a specified tolerance $\varepsilon$. The distance function may be a Kullback-Leibler divergence, Hellinger distance, or simply a mean-square error. The central idea of ABC is that, if the distance between modeled and reference data, $d(\mathcal{D}^{\prime}, \mathcal{D})$, is sufficiently small, then the parameter $\bm{c}$ is assumed to have been sampled from the posterior $P(\bm{c}\,|\,\mathcal{D})$.
	
In order to reduce the dimensionality of the data and, hence, the computational expense of ABC, it is common to replace the full reference data $\mathcal{D}$ with summary statistics $\mathcal{S}(\mathcal{D})$, such as the mean, standard deviation, or pdf of $\mathcal{D}$. The choice of summary statistic depends heavily on the problem and requires domain knowledge. However, a central assumption of ABC is that, if we use an appropriate summary statistic that depends on the choice of $\bm{c}$, the posteriors based on $\mathcal{S}$ and $\mathcal{D}$ will be equivalent, such that $P(\bm{c}\,|\,\mathcal{S}) = P(\bm{c}\,|\,\mathcal{D})$. 

Returning to the acceptance criterion in Eq.~\eqref{key} in the context of summary statistics, the distance between model and reference data can be replaced by a corresponding statistical distance for the modeled and reference summary statistics, denoted $\mathcal{S}'$ and $\mathcal{S}$, respectively. By applying the statistical distance function, $d(\mathcal{S}', \mathcal{S})$, introducing the acceptance threshold $\varepsilon$, and using a summary statistic instead of the full data, we finally obtain the ABC rejection sampling method given by Algorithm~\ref{alg:abc-rej}.

%algorithm-------------------------------
\begin{algorithm}[H]
	\caption{ABC rejection sampling algorithm}\label{alg:abc-rej}
	\begin{algorithmic}[1]
		\State Calculate reference summary statistic $\mathcal{S}$ from $\mathcal{D}$
		\State Sample $N$ sets of parameters $\bm{c}_i$ from prior distribution $\pi(\bm{c})$
		\For{each $\bm{c}_i$ }
		\State Calculate $\mathcal{D}^{\prime} = \mathcal{F}(\bm{c}_i)$ from model
		\State Calculate model summary statistic $\mathcal{S}'$ from $\mathcal{D}'$
		\State Calculate statistical distance $d(\mathcal{S}', \mathcal{S})$
		\If{$d(\mathcal{S}', \mathcal{S})\le\varepsilon$} 
		\State Accept $\bm{c}_i$
		\EndIf
		\EndFor
		\State Using all accepted $\bm{c}_i$ calculate posterior joint pdf
	\end{algorithmic}
\end{algorithm}
%algorithm-------------------------------
	
Although ABC is based on Bayes' theorem, it should be noted that, instead of determining the true posterior, ABC provides an approximation to the posterior distribution using a distance function and summary statistics~\cite{sunnaker_approximate_2013}. For appropriate summary statistics and as $\varepsilon \rightarrow 0$, the central assumption of ABC is that 
\begin{equation}
P(\bm{c}\,|\,\mathcal{D}) = \lim_{\varepsilon\rightarrow 0} P\left[\bm{c}\,|\, d(\mathcal{S}^{\prime}, \mathcal{S}) \le \varepsilon\right]\,.
\end{equation}
However, in practice, too small of an $\varepsilon$ is computationally impractical because it leads to too many rejections of sampled parameters $\bm{c}_i$ and an insufficiently converged posterior distribution. Conversely, relaxing the acceptance criterion (i.e., using a larger $\varepsilon$) and not selecting appropriate summary statistics can lead to bias in the final posterior distribution. 

The primary advantages of the ABC approach are the low cost relative to full Bayesian methods and the flexibility in parameter estimation for complex models. In particular, ABC does not require a likelihood function and enables parameter estimation for simulation-based models that can contain un-observable random quantities. A single set of parameters can be selected as the maximum \textit{a posteriori} probability (MAP) estimate, a mean value, or another characteristic statistic of the posterior distribution. Uncertainty can be quantified using confidence intervals or some other measure of the width of the posterior, including Monte Carlo sampling of parameter values from the posterior.

%===============================================
% Subsec: Monte Carlo Markov Chain
%===============================================
\subsection{ABC with Markov chain Monte Carlo sampling}
Although Algorithm~\ref{alg:abc-rej} is straightforward to implement, it can be computationally expensive and inefficient. The number of accepted parameters that form the posterior distribution are only a small fraction of the total number of sampled parameters, and the region (in parameter space) of accepted parameters rapidly shrinks as the number of parameters in the model increases. Thus, most of the sampled parameters and evaluations of summary statistics do not contribute to the posterior. However, this problem can be mitigated, and the sampling technique can be significantly improved, by using Markov chain Monte Carlo (MCMC) methods. 

The MCMC without likelihood method (or ABC-MCMC method) introduced by \citet{marjoram_markov_2003}, is based on the Metropolis-Hastings algorithm.  For an accepted parameter $\bm{c}_i$, the Metropolis-Hastings algorithm samples the next candidate parameter using the proposal $q(\bm{c}_i\rightarrow\bm{c}')$. If $d(\mathcal{S}^{\prime}, \mathcal{S})\le \varepsilon$, then the proposed parameters are accepted with probability 
\begin{equation}
h = \min\left[1, \frac{\pi(\bm{c}')q(\bm{c}_i\rightarrow \bm{c}')} {\pi(\bm{c_i})q(\bm{c'}\rightarrow \bm{c}_i)}\right]\,.
\end{equation}
Using the detailed balance condition, \citet{marjoram_markov_2003} demonstrated that this method yields a stationary posterior distribution, $P\left[\bm{c}\,|\, d(\mathcal{S}^{\prime}, \mathcal{S}) \le \varepsilon\right]$. An outline of the ABC-MCMC method is provided in Algorithm \ref{alg:MCMC}.

%algorithm-------------------------------
\begin{algorithm}[H]
	\caption{ABC-MCMC algorithm}\label{alg:MCMC}
	\begin{algorithmic}[1]
	    \State Calculate reference summary statistic $\mathcal{S}$ from $\mathcal{D}$
	    \State Define proposal covariance matrix $C$
		\State Start with initial accepted parameters $\bm{c}_0$
		\State $i:=0$
		\While{$i < N$}
		\State Sample $\bm{c}'$ from proposal $q(\bm{c}_i\rightarrow \bm{c}')=q(\bm{c'}\,|\,\bm{c}_i, C)$
		\State Calculate $\mathcal{D}^{\prime} = \mathcal{F}(\bm{c}')$ from model
		\State Calculate model summary statistic $\mathcal{S}'$ from $\mathcal{D}'$
		\State Calculate statistical distance $d(\mathcal{S}', \mathcal{S})$
		\If{$d(\mathcal{S}', \mathcal{S})\le \varepsilon$}
		\State Accept $\bm{c}'$ with probability
		$$h = \min\left[1, \frac{\pi(\bm{c}')q(\bm{c}_i\rightarrow \bm{c}')}{\pi(\bm{c}_i)q(\bm{c}'\rightarrow \bm{c}_i)}\right]$$
		\If{Accepted}
		\State Increment $i$ 
		\State Set $\bm{c}_i = \bm{c}'$ 
		\EndIf
		\EndIf
		\EndWhile
		\State Using all accepted $\bm{c}_i$ calculate posterior joint pdf
	\end{algorithmic}
\end{algorithm}
%algorithm-------------------------------

In the present demonstration of ABC-MCMC for SGS model parameter estimation, we choose the proposal distribution to be Gaussian with the current parameter $\bm{c}_i$ as the mean value and the covariance matrix $C$ as the width and orientation (in parameter space); i.e., $q(\bm{c}_i\rightarrow\bm{c}') = q(\bm{c'}\,|\,\bm{c}_i, C)$. For a Gaussian proposal, $q(\bm{c}_i\rightarrow \bm{c}') = q(\bm{c'}\rightarrow \bm{c}_i)$ and $h$ depends only on the prior. If the prior is uniform,  then $\pi(\bm{c}_i) = \pi(\bm{c}')$, and the Gaussian proposal leads  to $h = 1$ for any $\bm{c}'$. As a result, the algorithm reduces to a rejection method with correlated outputs~\cite{marjoram_markov_2003}. Here, we update the covariance $C$ as the ABC-MCMC rejection sampling algorithm proceeds, leading to an adaptive proposal. 

%===============================================
% Sec: Adaptive Proposal
%===============================================
\subsection{Adaptive proposal\label{sec:method_adaptive}}
The choice of proposal distribution $q$ has a significant impact on the rate of convergence of the Metropolis-Hastings algorithm. For many problems, a variable proposal with an adaptive size and orientation (in parameter space) provides faster convergence, and we thus implement an adaptive proposal in the ABC-MCMC approach. 

Here we follow the adaptive proposal procedure outlined by \citet{haario_adaptive_2001}, where the covariance $C_i$ after the $i$th accepted parameters $\bm{c}_i$ in the Gaussian proposal $q(\bm{c}'\,|\,\bm{c}_i,C_i)$ is updated during the process using all previous steps of the chain as
\begin{equation}
C_i = 
\begin{cases} 
s_n\mathrm{var}I, & \mbox{if } i < k\\ 
s_n\mathrm{cov}(\bm{c}_0, \dots, \bm{c}_i), & \mbox{if } i\ge k 
\end{cases},
\end{equation}
where $k>0$ is the length of the initial period without adaptation and $s_n$ is a constant that depends on the dimensionality of the parameter space, $n$, as $s_n = (2.4)^2/n$. This algorithm does not require substantial additional computational cost, since the covariance $\mathrm{cov}_{i}\equiv \mathrm{cov}(\bm{c}_0, \dots, \bm{c}_i)$ can be calculated using the recursive formula
\begin{equation}
\mathrm{cov}_{i} = \left(\frac{i-1}{i}\right)\mathrm{cov}_{i-1} + \left(\frac{1}{i+1}\right)\left(\boldsymbol{\mu}_{i-1} - \bm{c}_i \right)\left(\boldsymbol{\mu}_{i-1} - \bm{c}_i \right)^T\,,
\end{equation}
where $\boldsymbol{\mu}_i = [1/(i+1)]\sum_{k=0}^i \bm{c}_k$ is the average of all previous accepted parameter values, which also satisfies the recursive formula
\begin{equation}
\boldsymbol{\mu}_i = \boldsymbol{\mu}_{i-1} - \left(\frac{i}{i + 1}\right) \left(\boldsymbol{\mu}_{i-1} - \bm{c}_i \right)\,.
\end{equation} 
Although the adaptive Metropolis algorithm is no longer Markovian, \citet{haario_adaptive_2001} have shown that it has the correct ergodic properties and the accuracy of the algorithm is close to the accuracy of the standard Metropolis-Hastings algorithm given a properly chosen proposal.

%===============================================
% Sec: Calibration step
%===============================================
\subsection{Calibration step}
The main advantage of the ABC-MCMC method is the high rate of acceptance, since we start from an accepted parameter and stay in the acceptance region. However, for parameter spaces with large dimensions $n$, the acceptance region becomes quite small relative to the total size of the parameter space. As a result, determining the initial accepted parameters $\bm{c}_0$ can itself require many iterations. 

The other primary challenge in implementing ABC-MCMC is the same as for ABC rejection approaches more generally; namely, the fixed acceptance threshold $\varepsilon$ must be defined \emph{a priori}, and indeed before the entire Markov chain is run. The choice of $\varepsilon$ is important, since too large a tolerance results in a chain that is dominated by the prior. On the other hand, too small a tolerance leads to a very small acceptance rate and also increases the initialization cost. 

%algorithm-------------------------------
\begin{algorithm}[H]
	\caption{ABC-IMCMC algorithm with an initial calibration step and an adaptive proposal}\label{alg:IMCMC}
	\begin{algorithmic}[1]
	    \State Calculate reference summary statistic $\mathcal{S}$ from $\mathcal{D}$
		\Procedure{Calibration step}{$N_\mathrm{c}$, $r$}
		\State Sample $N_\mathrm{c}$ parameters $\bm{c}_i$ from prior distribution $\pi(\bm{c})$
		\For{each $\bm{c}_i$ }
		\State Calculate $\mathcal{D}^{\prime} = \mathcal{F}(\bm{c}_i)$ from model
		\State Calculate model summary statistic $\mathcal{S}'$ from $\mathcal{D}'$
		\State Calculate statistical distance $d_i(\mathcal{S}', \mathcal{S})$
		\EndFor
		\State Using all $d_i(\mathcal{S}', \mathcal{S})$ calculate distribution $P(d)$
		\State Define tolerance $\varepsilon$ such that $P(d\le\varepsilon)=r$
		\State Randomly choose $\bm{c}_0$ from $\bm{c}_i$ parameters with $d\le\varepsilon$
		\State Adjust prior based on variance of parameters with $d\le \varepsilon$
		\State Calculate covariance $C_0$ from parameters with $d\le\varepsilon$
		\EndProcedure
		\Procedure{MCMC without likelihood}{$\bm{c}_0$, $\varepsilon$, $C_0$, $k$, $N$}
		\State Start with accepted parameters $\bm{c}_0$ and covariance $C_0$ 
		\State $i:=0$
		\While{$i < N$}
		\State Sample $\bm{c}'$ from proposal $q(\bm{c}_i\rightarrow \bm{c}')=q(\bm{c'}\,|\,\bm{c}_i, C_i)$
		\State Calculate $\mathcal{D}^{\prime} = \mathcal{F}(\bm{c}')$ from model
		\State Calculate model summary statistic $\mathcal{S}'$ from $\mathcal{D}'$
		\State Calculate statistical distance $d(\mathcal{S}', \mathcal{S})$
		\If{$d(\mathcal{S}', \mathcal{S})\le \varepsilon$}
		\State Accept $\bm{c}'$ with probability
		$$h = \min\left[1, \frac{\pi(\bm{c}')q(\bm{c}_i\rightarrow \bm{c}')}{\pi(\bm{c}_i)q(\bm{c}'\rightarrow \bm{c}_i)}\right]$$
		\If{Accepted}
		\State Increment $i$ 
		\State Set $\bm{c}_i = \bm{c}'$
		\State Update proposal covariance $C_i$ as
		\[C_i = 
        \begin{cases} 
        s_n\mathrm{var}I, & \mbox{if } i < k\\ 
        s_n\mathrm{cov}(\bm{c}_0, \dots, \bm{c}_i), & \mbox{if } i\ge k 
        \end{cases}\]
        \EndIf
		\EndIf
		\EndWhile
		\State Using all accepted $\bm{c}_i$ calculate posterior joint pdf
		\EndProcedure
	\end{algorithmic}
\end{algorithm}
% algorithm -------------------------------

%\begin{figure}[t!]
%	\centering
%	\includegraphics[scale=0.85]{algorithms-figure3}
%	\caption{Schematic representation of the ABC-IMCMC algorithm with a calibration step and an adaptive proposal. The schematic corresponds to Algorithm \ref{alg:IMCMC}.}
%	\label{fig:IMCMC}
%\end{figure}
%figure-------------------------------

Proposal parameters such as the initial variance of the proposal $q$ must also be defined \textit{a priori}. The fraction of accepted parameters in ABC-MCMC depends on the variance in the proposal, such that a smaller variance leads to a larger number of accepted parameters, but more iterations are required to obtain a statistically converged posterior distribution. 

To address these issues, \citet{wegmann_efficient_2009} suggested an ``improved'' version of the MCMC algorithm, denoted IMCMC, that has an initial calibration step. In this step, a series of $N_\mathrm{c}$ simulations are performed, where the parameters are uniformly sampled from their prior $\pi(\bm{c})$ to obtain an approximate probability distribution, $P(d)$, of the distances $d(\mathcal{S}^\prime,\mathcal{S})$ over the entire parameter space. Using this calibration step, we can define a threshold distance $\varepsilon$ such that $P(d\leq\varepsilon)=r$, where $r$ is a desired ratio of accepted simulation parameters. We can then use any simulation for which $d \leq \varepsilon$ as a starting point for the Markov chain. These $N_\mathrm{c}$ calibration simulations are also used to adjust the transition kernel $q(\bm{c}_i\rightarrow \bm{c}')$. In our case, the initial $q$ is a Gaussian distribution with standard deviation equal to the standard deviation of the retained parameters from the calibration step. The resulting Algorithm \ref{alg:IMCMC} is based on IMCMC with the calibration step from~\cite{wegmann_efficient_2009}, and is the complete form of ABC-IMCMC used here to solve the SGS model parameter estimation inverse problem. 

\section{Setup of ABC-IMCMC for Model Parameter Estimation\label{sec:details}}
We demonstrate the use of ABC-IMCMC to determine the parameters $\bm{c}$ in Eq.~\eqref{eq:closure} for three- and four-term versions of the nonlinear SGS model (corresponding to parameter space dimensions $n=3$ and 4, respectively). For both of these models, we perform ABC-IMCMC parameter estimation for distance functions based only on summary statistics of $\sigma_{ij}$, based only on summary statistics of $\mathcal{P}$, and based on a combination of summary statistics of $\sigma_{ij}$ and $\mathcal{P}$. As such, we will estimate three sets of parameters $\bm{c}$ for both the three- and four-parameter models, for a total of six different tests of ABC-IMCMC. The corresponding joint posterior distributions obtained in these tests are denoted $P^{(n)}_{\mathcal{D}}(\bm{c})$, where $n$ denotes the number of parameters in the model (either 3 or 4), and $\mathcal{D}$ denotes the underlying reference data used in the ABC-IMCMC parameter estimation (i.e., $\sigma$ for SGS stress tensor reference data, $\mathcal{P}$ for SGS production reference data, or $\sigma\mathcal{P}$ for combined SGS stress and production reference data).

In the following, we describe the specific setup of the ABC-IMCMC procedure (summarized in general form in Algorithm \ref{alg:IMCMC}) used for these tests, including the choices of reference data, summary statistics, distance functions, and priors. It is emphasized, however, that the ABC-IMCMC procedure outlined in the previous section is completely general and can be applied to different model forms using different configurations (e.g., different reference data, summary statistics, distance functions, and priors). As such, the current ABC-IMCMC configuration for SGS model parameter estimation should be taken as demonstrative of the power and limitations of the approach, rather than as a comprehensive description of the ideal formulation of ABC-IMCMC, either in terms of computational efficiency or accuracy. In particular, through the use of different types of reference data, we seek to demonstrate the impact of the choice of reference data on the estimated parameter distributions. Similarly, through the use of two different model forms, we show the inherent limitations imposed by the physical accuracy of the model; namely, an imperfect model cannot be made to perfectly agree with reference data simply by calibrating a finite number of model parameters.  

\subsection{Reference and model data}
%figure-------------------------------
\begin{figure}[t!]
	\centering
	\includegraphics{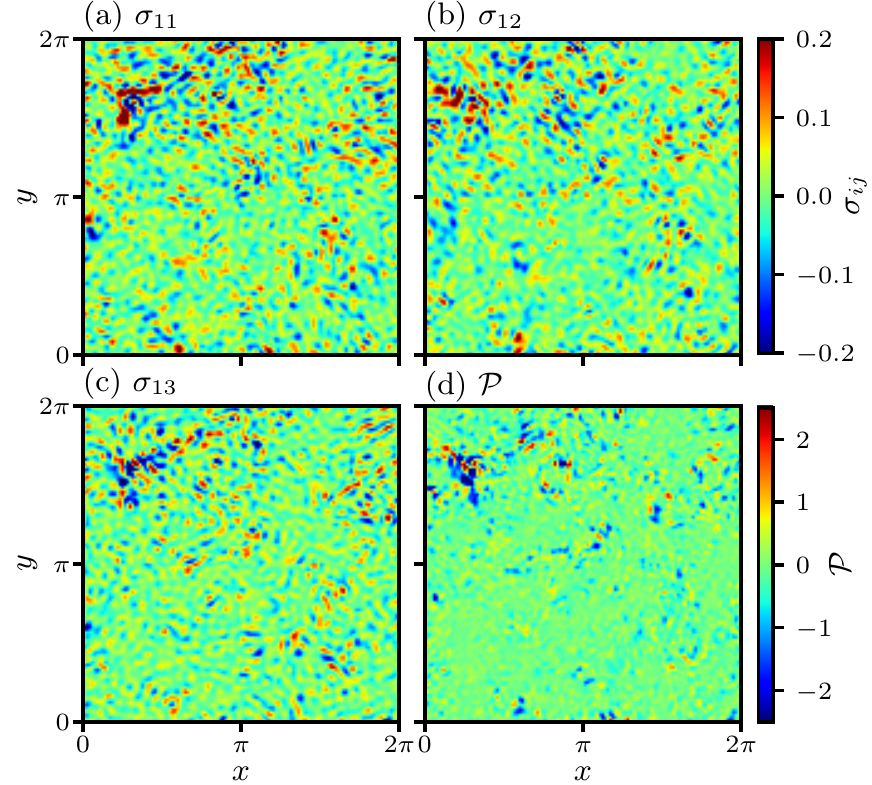}
	\caption{Two-dimensional fields of HIT DNS \cite{li2008} reference data used in the present ABC-IMCMC demonstration, showing (a) $\mathcal{D}_{11}=\sigma_{11}$, (b) $\mathcal{D}_{12}=\sigma_{12}$, (c) $\mathcal{D}_{13}=\sigma_{13}$, and (d) $\mathcal{D}_\mathcal{P} = \sigma_{ij} \widetilde{S}_{ij}$.}
	\label{fig:fields}
\end{figure}
%figure-------------------------------
The reference data are obtained from the Johns Hopkins Turbulence Database~\cite{li2008} by sampling every fourth point in a triply periodic $1024^3$ cube of DNS data for homogeneous isotropic turbulence (HIT) at $Re_\lambda = 433$. The resulting $256^3$ cube of sub-sampled data is then filtered at wavenumber $k_{\Delta} = 30$ using a spectrally sharp filter to obtain fields of the resolved-scale velocity vector $\widetilde{u}_i$ and the second-order velocity product tensor $\widetilde{u_i u_j}$. From these fields, we obtain the reference data for the present tests, given as $\mathcal{D}_{ij}=\sigma_{ij}$ for tests using the deviatoric SGS stress tensor as reference data and $\mathcal{D}_\mathcal{P}=\sigma_{ij} \widetilde{S}_{ij}$ for tests using the SGS production as reference data. Two-dimensional fields of reference data $\sigma_{11}$, $\sigma_{12}$, $\sigma_{13}$, and $\mathcal{P}$ from the DNS are shown in Figure \ref{fig:fields}. 

The corresponding model data are obtained as $\mathcal{D}'_{ij}=\mathcal{F}_{ij}(\bm{c})$ and $\mathcal{D}'_\mathcal{P}=\mathcal{F}_{ij}(\bm{c})\widetilde{S}_{ij}$, where $\mathcal{F}_{ij}(\bm{c})$ represents the model approximation for $\sigma_{ij}$ for a given choice of $\bm{c}$ from either the three ($n=3$) or four ($n=4$) parameter SGS stress closures given by Eq.~\eqref{eq:closure}. The resolved scale strain and rotation rate tensors, $\widetilde{S}_{ij}$ and $\widetilde{R}_{ij}$, respectively, required for the calculation of $\mathcal{F}_{ij}(\bm{c})$ are obtained from the DNS data.

\subsection{Summary statistics}
\begin{figure}[t!]
	\centering
	\includegraphics{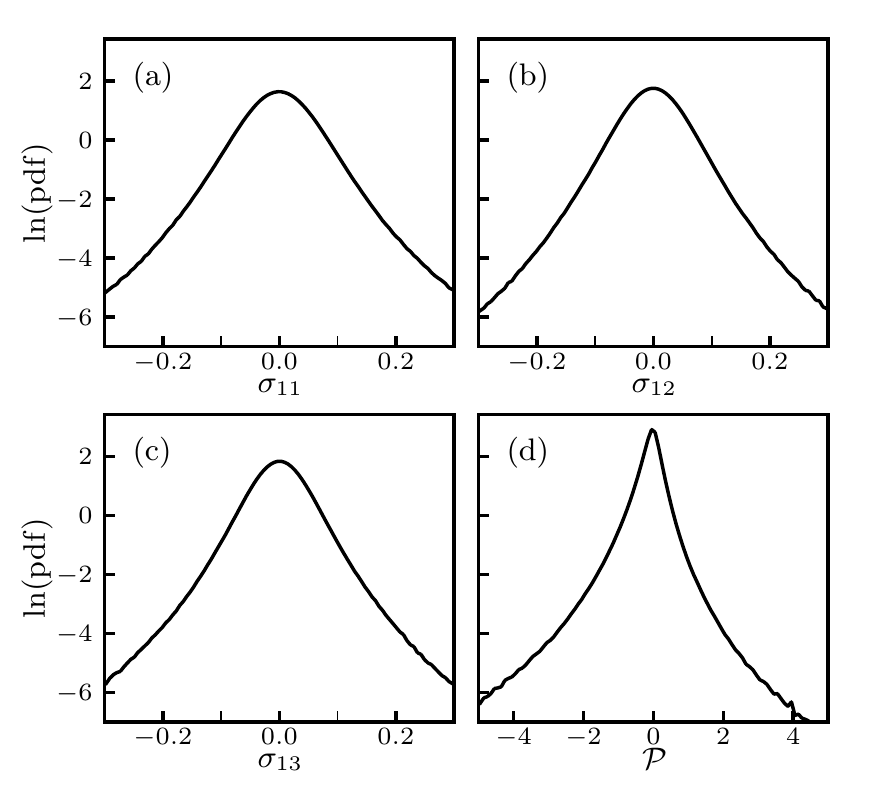}
	\caption{Summary statistics from the HIT DNS \cite{li2008} reference data used in the present ABC-IMCMC demonstration, showing pdfs of (a) $\mathcal{D}_{11}=\sigma_{11}$, (b) $\mathcal{D}_{12}=\sigma_{12}$, (c) $\mathcal{D}_{13}=\sigma_{13}$, and (d) $\mathcal{D}_\mathcal{P} = \sigma_{ij} \widetilde{S}_{ij}$. These summary statistics are denoted $\mathcal{S}_{11}$, $\mathcal{S}_{12}$, $\mathcal{S}_{13}$, and $\mathcal{S}_\mathcal{P}$, respectively.}
	\label{fig:summary}
\end{figure}

All summary statistics in the present tests are based on pdfs of the reference and model data. For the deviatoric stresses, the summary statistics (i.e., pdfs) are denoted $\mathcal{S}_{ij}$ and $\mathcal{S}^{\prime}_{ij}$ for the reference ($\mathcal{D}_{ij}$) and model ($\mathcal{D}'_{ij}$) data, respectively, where there are separate pdfs for each $(i,j)$ component of the stress tensor. We also consider pdfs of the production $\mathcal{P}$, denoted $\mathcal{S}_\mathcal{P}$ and $\mathcal{S}^{\prime}_\mathcal{P}$ for $\mathcal{D}_\mathcal{P}$ and $\mathcal{D}'_\mathcal{P}$, respectively. The flexibility of ABC-IMCMC also allows us to combine summary statistics, and in the following we will present parameter estimation results using both the stress and production summary statistics simultaneously. The reference data summary statistics for $\mathcal{S}_{11}$, $\mathcal{S}_{12}$, $\mathcal{S}_{13}$, and $\mathcal{S}_\mathcal{P}$ are shown in Figure \ref{fig:summary}.

It should be noted that, since we do not expect either of the SGS closures examined here to exactly represent the flow physics, a degree of uncertainty must be present in the model summary statistics $\mathcal{S}^{\prime}_{ij}$ and $\mathcal{S}^{\prime}_\mathcal{P}$ to avoid over-fitting. To introduce this additional uncertainty, we randomly choose $10^5$ data points out of the filtered $256^3$ cube of DNS reference data for each ABC-IMCMC iteration to calculate $\mathcal{S}^{\prime}_{ij}$ and $\mathcal{S}^{\prime}_\mathcal{P}$. This approach has the dual benefit of making our model stochastic while also reducing the total number of computations in the ABC-IMCMC procedure. 

%figure-------------------------------
%figure-------------------------------

\subsection{Statistical distances and acceptance criteria}
The statistical distances between the modeled and reference summary statistics can be calculated many different ways, and here we use the Mean Square Error (MSE) of the logarithms of the pdfs.  We use the logarithms of the pdfs in the present tests to emphasize the importance of the pdf ``tails'', which correspond to lower probabilities of extreme values, as shown in Figure \ref{fig:summary}.

For the SGS stress summary statistics, we calculate the distance, denoted $d_\sigma(\mathcal{S}_{ij}', \mathcal{S}_{ij})$, as 
\begin{equation}
d_\sigma(\mathcal{S}_{ij}', \mathcal{S}_{ij}) = \sum_{i,j\,|\, j\geq i}\overline{\left(\ln\mathcal{S}_{ij}' - \ln\mathcal{S}_{ij}\right)^2}\,, 
\end{equation}
where the summation is over all $i$ and $j$ such that $j\geq i$, giving six independent terms in the summation; this summation approach is necessary since the stress tensors are symmetric and the terms with $i\neq j$ should not be double-counted in the combined distance metric. The average $\overline{(\cdot)}$ is performed over all bins in the pdfs. Similarly, for the SGS production summary statistics, we calculate the distance $d_\mathcal{P}(\mathcal{S}_\mathcal{P}', \mathcal{S}_\mathcal{P})$ as 
\begin{equation}
d_\mathcal{P}(\mathcal{S}_\mathcal{P}', \mathcal{S}_\mathcal{P}) = \overline{\left(\ln\mathcal{S}_\mathcal{P}' - \ln\mathcal{S}_\mathcal{P}\right)^2}\,.
\end{equation}
For both the stress- and production-based distance functions, the acceptance criteria are simply $d_{\sigma}\leq \varepsilon$ and $d_\mathcal{P}\leq \varepsilon$, respectively. The specific values of $\varepsilon$ are determined during an initial calibration step, as outlined in the next section.

Finally, when using a combined stress and production acceptance criterion, we define a new distance function $d_{\sigma\mathcal{P}}$ as
\begin{equation}\label{eq:dist_combined}
d_{\sigma\mathcal{P}}=\alpha d_\sigma + \beta d_\mathcal{P}\,,
\end{equation}
%figure-------------------------------
where $\alpha$ and $\beta$ are weighting coefficients that can be changed based on the desired importance of different summary statistic components. For the present ABC-IMCMC demonstration, we use $\alpha=1$ and $\beta=3$, although different values can be used to emphasize greater model correspondence to either the $\sigma_{ij}$ or $\mathcal{P}$ pdfs. Once again, the acceptance criterion is $d_{\sigma\mathcal{P}}\leq \varepsilon$, where $\varepsilon$ is determined during the initial calibration step.

\subsection{Calibration step and priors}
As described in Section \ref{sec:method}, we perform an initial calibration step to choose different appropriate values of $\varepsilon$ for each of the three types of distance function considered here. We also use this step to determine the initial choices of parameter $\bm{c}_0$ and proposal covariance $C_0$. 

For all choices of distance function, we used $N_\mathrm{c}=10^3$ and $10^4$ samples of $\bm{c}$, taken from uniform priors, in the calibration step for the three- and four-parameter models, respectively. For the calibration based on $d_\sigma$, the uniform priors were bounded by $c_1 \in[-0.3, 0.0]$, $c_2\in[-0.5, 0.5]$, $c_3\in[-0.2, 0.2]$, and  $c_4\in[-0.2, 0.2]$, where the $c_4$ dimension is not used for the three-parameter model tests. For the calibrations based on $d_\mathcal{P}$ and $d_{\sigma\mathcal{P}}$, the priors were bounded by $c_1 \in[-0.3, 0.0]$, $c_2\in[-0.5, 0.5]$, $c_3\in[-0.5, 0.2]$ and  $c_4\in[-0.2, 0.5]$, where again the $c_4$ dimension was only used for the four-parameter model tests. 

%figure-------------------------------
\begin{figure}[t!]
	\centering
	\includegraphics{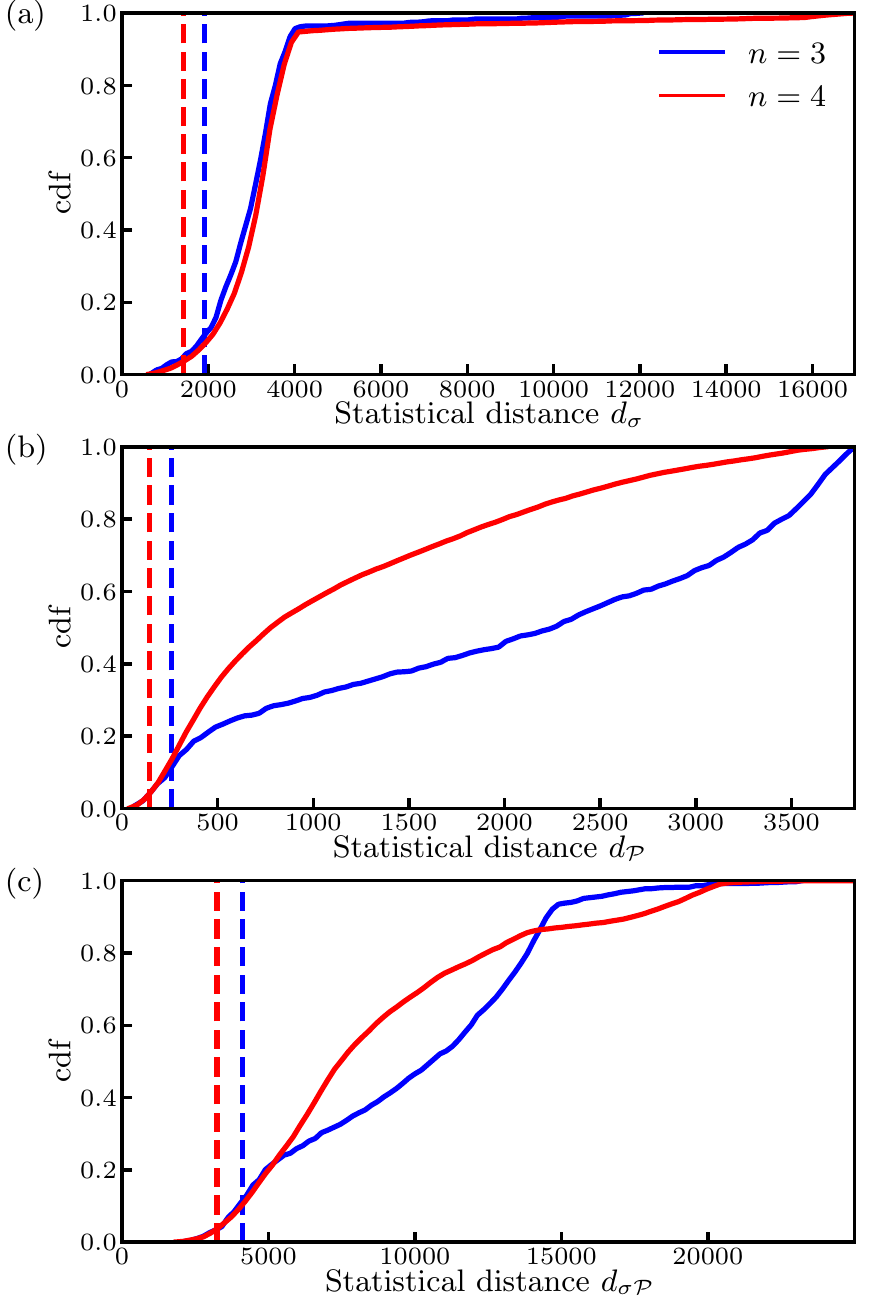}
	\caption{Cumulative distribution functions (cdfs) of the distances (a) $d_\sigma$, (b) $d_\mathcal{P}$, and (c) $d_{\sigma\mathcal{P}}$ for the three- ($n=3$) and four- ($n=4$) parameter SGS models from Eq.~\eqref{eq:closure} (denoted by blue and red lines, respectively). Vertical dashed lines show values of $\varepsilon_\mathcal{D}^{(n)}$ for each distance function and model type (blue and red dashed lines for the three- and four-parameter models, respectively). Values of $\varepsilon_\mathcal{D}^{(n)}$ were chosen based on an acceptance ratio of $r=10\%$ for the three-parameter model and $r=3\%$ for the four-parameter model.}
	\label{fig:calibrate}
\end{figure}
%figure-------------------------------

From the calibration steps, $\varepsilon$ was chosen for each distance to correspond to an acceptance percentage of $r=10\%$ for the three-parameter model and $r=3\%$ for the four-parameter model. The cumulative distribution functions for each of the calibration steps are shown in Figure \ref{fig:calibrate}, with the values of $\varepsilon$ resulting from these tests indicated by dashed lines. There are six resulting values of $\varepsilon$, denoted $\varepsilon_\mathcal{D}^{(n)}$, corresponding to each of the six posterior distributions $P_\mathcal{D}^{(n)}(\bm{c})$ that will be computed in the present demonstration of ABC-IMCMC. 

%%%%%%%%%%%%%%%%%%%%%%%%%%%%%%%%
% Sec: Results
%%%%%%%%%%%%%%%%%%%%%%%%%%%%%%%%
\section{Results\label{sec:results}}
Here we apply ABC-IMCMC to estimate posterior distributions of unknown parameters for the three- and four-parameter versions of the SGS stress model given by Eq.~\eqref{eq:closure}. The posteriors are obtained separately for each model using three different types of model and reference data: (\emph{i}) SGS stresses $\sigma_{ij}$, giving the posterior $P^{(n)}_{\sigma}(\bm{c})$, where $n$ is the number of parameters in the model; (\emph{ii}) SGS production $\mathcal{P}$, giving $P^{(n)}_{\mathcal{P}}(\bm{c})$; and (\emph{iii}) a combination of $\sigma_{ij}$ and $\mathcal{P}$, giving $P^{(n)}_{\sigma\mathcal{P}}(\bm{c})$. The detailed configuration for each of these tests is described in the previous section. 

In the following, we first present the posterior distributions resulting from ABC-IMCMC for both three- and four-parameter models and all three types of reference data. From the posteriors, we obtain the MAP values of the model parameters, which are then used in \emph{a priori} tests to show that true summary statistics can be recovered by the models using the estimated parameters. We also present results from \emph{a posteriori} tests showing that models using the estimated parameters can be stably integrated in forward LES runs. Uncertainties in the estimated parameters are naturally given by the posterior distributions resulting from the ABC-IMCMC approach, and we estimate the impact of these uncertainties in the \emph{a posteriori} tests by performing the LES for many different parameter samples from the posteriors, and then weighting the results by the posterior value for that sample.

%===============================================
% Subsec: Posteriors
%===============================================
\subsection{Model parameter posterior distributions}
\begin{figure*}[t!]
	\centering\includegraphics[width=\textwidth]{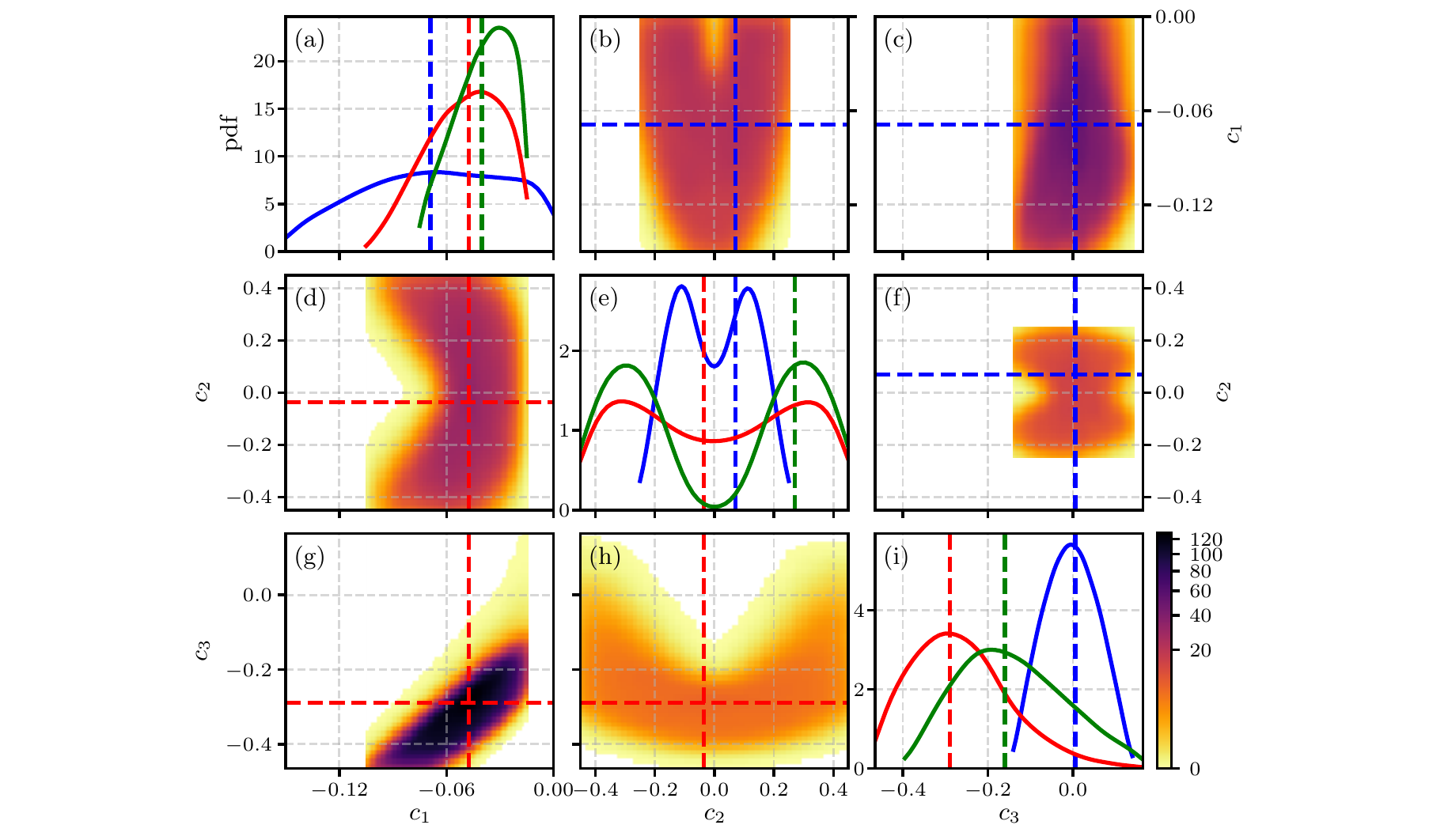}\vspace{-0.2in}
	\caption{Posterior distributions of accepted parameters $c_1$, $c_2$, and $c_3$ for the three-term SGS model from Eq.~\eqref{eq:closure}. The diagonal subplots show marginal distributions $M^{(3)}_\mathcal{\sigma}(c_i)$ (blue lines), $M^{(3)}_\mathcal{\mathcal{P}}(c_i)$ (red lines), and $M^{(3)}_\mathcal{\sigma\mathcal{P}}(c_i)$ (green lines) for (a) $i=1$, (e) $i=2$, and (i) $i=3$. The over-diagonal subplots (b,c,f) show $P^{(3)}_{\sigma}(c_i,c_j\,|\, c_k^\mathrm{MAP})$ and the under-diagonal subplots (d,g,h) show $P^{(3)}_{\mathcal{P}}(c_i,c_j\,|\, c_k^\mathrm{MAP})$. Dashed lines in each subplot show the MAP parameter estimates $c_i^\mathrm{MAP}$ in Table \ref{tab:MAP} for stress (red dashed lines), production (blue dashed lines), and combined stress and production (green dashed lines) data.}
	\label{fig:3_params}
\end{figure*}

\begin{figure*}[t!]
	\centering\includegraphics[width=\textwidth]{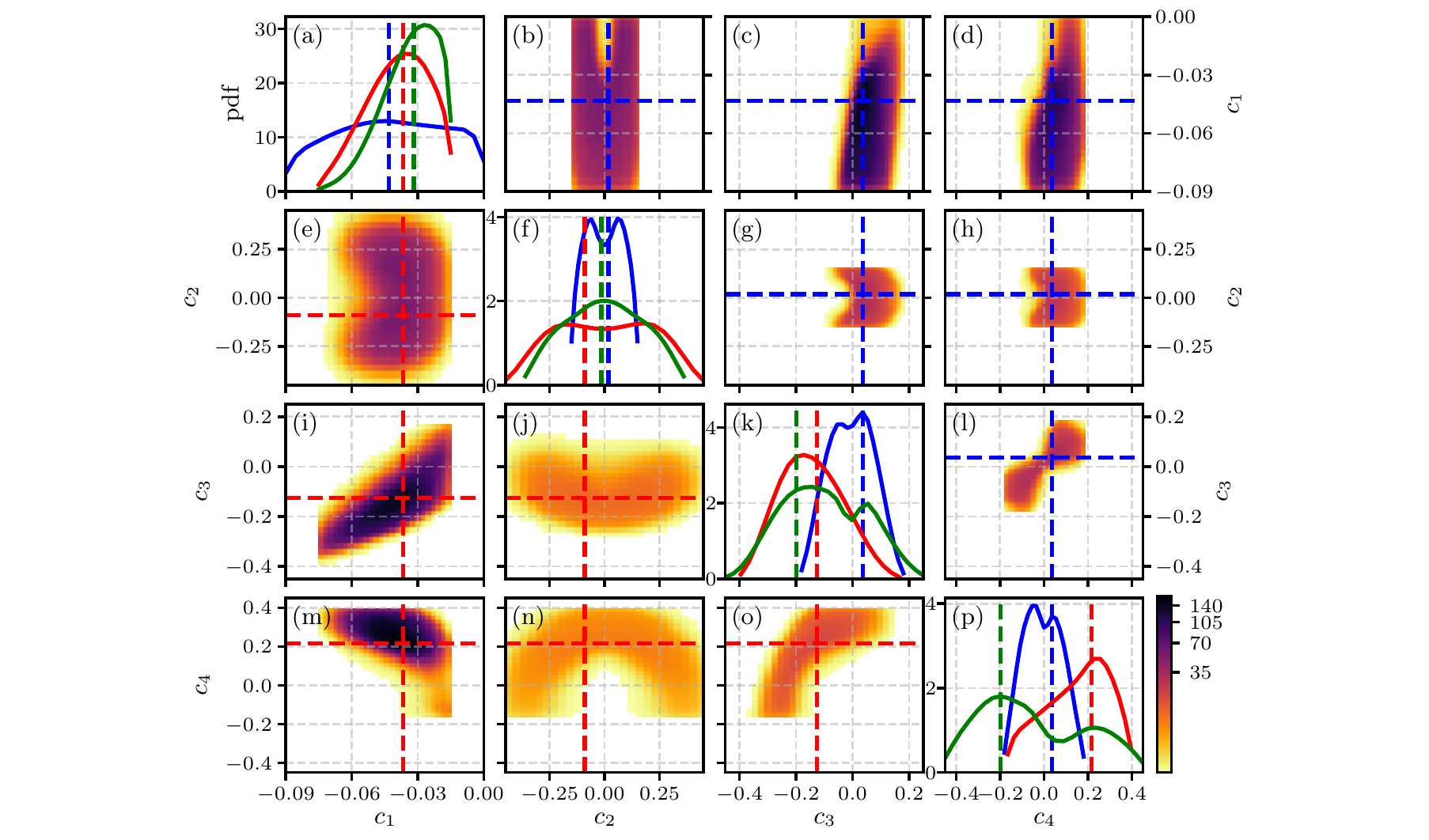}\vspace{-0.2in}
	\caption{Posterior distributions of accepted parameters $c_1$, $c_2$, $c_3$, and $c_4$ for the four-term SGS model from Eq.~\eqref{eq:closure}. The diagonal subplots show marginal distributions $M^{(4)}_\mathcal{\sigma}(c_i)$ (blue lines), $M^{(4)}_\mathcal{\mathcal{P}}(c_i)$ (red lines), and $M^{(4)}_\mathcal{\sigma\mathcal{P}}(c_i)$ (green lines) for (a) $i=1$, (f) $i=2$, (k) $i=3$, and (p) $i=4$. The over-diagonal subplots (b,c,d,g,h,l) show $P^{(4)}_{\sigma}(c_i,c_j\,|\, c_k^\mathrm{MAP},c_l^\mathrm{MAP})$ and the under-diagonal subplots (e,i,j,m,n,o) show $P^{(4)}_{\mathcal{P}}(c_i,c_j\,|\, c_k^\mathrm{MAP},c_l^\mathrm{MAP})$. Dashed lines in each subplot show the MAP parameter estimates $c_i^\mathrm{MAP}$ in Table \ref{tab:MAP} for stress (red dashed lines), production (blue dashed lines), and combined stress and production (green dashed lines) data.}
	\label{fig:4_params}
\end{figure*}

Figures~\ref{fig:3_params} and \ref{fig:4_params} show posterior distributions obtained from ABC-IMCMC for, respectively, three- and four-parameter SGS closure models. After the calibration step described in the previous section, these posteriors were generated by running parallel Markov chains with a total of roughly $N = 10^5$ and $10^7$ accepted parameters for the three- and four-parameter models, respectively. There were 6 parallel chains with acceptance rates between 48\% and 62\% for the three-parameter model and 24 parallel chains with acceptance rates between 31\% and 52\% for the four-parameter model.

In order to visualize the resulting three- and four-dimensional posteriors, in Figures~\ref{fig:3_params} and \ref{fig:4_params} we show one-dimensional marginal pdfs for each parameter on the diagonal subplots for each of the three types of statistical distances. These marginal distributions are denoted here $M^{(n)}_\mathcal{D}(c_i)$, where $c_i$ is the parameter value in the pdf after marginalization (i.e., summation) over all other values of $c_j$, with $i\neq j$. In both Figures~\ref{fig:3_params} and \ref{fig:4_params}, two-dimensional slices (in parameter space) from the posterior joint pdfs are shown at the MAP values of all other parameters, with $P^{(3)}_{\sigma}(c_i,c_j\,|\, c_k^\mathrm{MAP})$ and $P^{(4)}_{\sigma}(c_i,c_j\,|\, c_k^\mathrm{MAP},c_l^\mathrm{MAP})$ on the over-diagonal of each figure and $P^{(3)}_{\mathcal{P}}(c_i,c_j\,|\, c_k^\mathrm{MAP})$ and $P^{(4)}_{\mathcal{P}}(c_i,c_j\,|\, c_k^\mathrm{MAP},c_l^\mathrm{MAP})$ on the under-diagonal. Here, $i$, $j$, $k$, and $l$ each take on values between 1 and 4 (or only 3 for the three-parameter model), with none of the indices equal. All posterior pdfs are calculated using kernel density estimation with a Gaussian kernel and bandwidth defined by Scott's Rule~\cite{scott2015multivariate, silverman1986density}. A summary of the MAP parameter values for each test are provided in Table \ref{tab:MAP}.

%figure-------------------------------
%figure-------------------------------

%figure-------------------------------
%figure-------------------------------

For the three-parameter model, Figure~\ref{fig:3_params} and Table \ref{tab:MAP} show that ABC-IMCMC provides a MAP value of $c_1^\mathrm{MAP}= -0.069$ when using SGS stress data, and $c_1^\mathrm{MAP}=-0.047$ when using SGS production data. The 1D marginal posteriors $M^{(3)}_\mathcal{D}(c_2)$ are bimodal for all types of data, with MAP values of $c_2^\mathrm{MAP}=0.070$ for stress data and $c_2^\mathrm{MAP}=-0.036$ for production data. The bimodality of $M^{(3)}_\mathcal{D}(c_2)$ indicates that the sign of $c_2^\mathrm{MAP}$ is of little importance compared to the magnitude. The 2D joint pdfs $P^{(3)}_{\mathcal{D}}(c_2,c_j\,|\, c_k^\mathrm{MAP})$ are also symmetric with respect to $c_2$ for all types of data $\mathcal{D}$. 

The greatest dependence on the type of data for the three-parameter model is observed for $c_3$, where $c_3^\mathrm{MAP}$ in Table \ref{tab:MAP} is close to zero for stress data and $c_3^\mathrm{MAP}=-0.29$ for production data. The former result indicates that the $c_3$ term is of negligible importance compared to the other two terms when attempting to predict the stress pdfs. Moreover, the joint pdf of $P^{(3)}_{\mathcal{P}}(c_1,c_3 \,|\, c_2^\mathrm{MAP})$ in Figure~\ref{fig:3_params}(g) shows that these two parameters are correlated when using production data, but there is no similarly strong correlation when using stress data [i.e., in Figure~\ref{fig:3_params}(c)] . 

When using both the stress and production data, Figure~\ref{fig:3_params} and Table \ref{tab:MAP} show that the 1D marginal pdfs and associated MAP values are generally a combination of the stress and production data results, with MAP values of $c_1^\mathrm{MAP}=-0.040$, $c_2^\mathrm{MAP}=0.27$, and $c_3^\mathrm{MAP}=-0.16$. In general, however, the combined results are most similar to results obtained using production data alone. 

%table-------------------------------
\begin{table}[t!]
\centering
\begin{tabular}{ccc|cccccc}
\hline\hline
 & & & \multicolumn{4}{c}{MAP parameter estimates, $c_i^\mathrm{MAP}$}\\ 
Data            & $n$   & Posterior                         & $c_1$     & $c_2$     & $c_3$     & $c_4$ \\ \hline
$\sigma_{ij}$   & 3     & $P^{(3)}_{\sigma}(\bm{c})$        & -0.069    & 0.070     & 0.0056    & --    \\
$\sigma_{ij}$   & 4     & $P^{(4)}_{\sigma}(\bm{c})$        & -0.043    & 0.018     & 0.036     & 0.036 \\ \hline
$\mathcal{P}$   & 3     & $P^{(3)}_{\mathcal{P}}(\bm{c})$   & -0.047    & -0.036    & -0.29     & --    \\ 
$\mathcal{P}$   & 4     & $P^{(4)}_{\mathcal{P}}(\bm{c})$   & -0.037    & -0.090    & -0.13     & 0.22 \\ \hline
$\sigma_{ij}$, $\mathcal{P}$ & 3 & $P^{(3)}_{\sigma\mathcal{P}}(\bm{c})$ & -0.040 & 0.27   & -0.16 & -- \\ 
$\sigma_{ij}$, $\mathcal{P}$ & 4 & $P^{(4)}_{\sigma\mathcal{P}}(\bm{c})$ & -0.032 & -0.014 & -0.20 &  -0.20\\\hline\hline
\end{tabular}
\caption{MAP parameter estimates for the three ($n=3$) and four ($n=4$) parameter SGS models in Eq.~\eqref{eq:closure} from posterior joint probability density functions calculated using ABC-IMCMC for different choices of data.}
\label{tab:MAP}
\end{table}
%table-------------------------------

%figure-------------------------------
\begin{figure*}[b!]
	\centering\includegraphics[width=\linewidth]{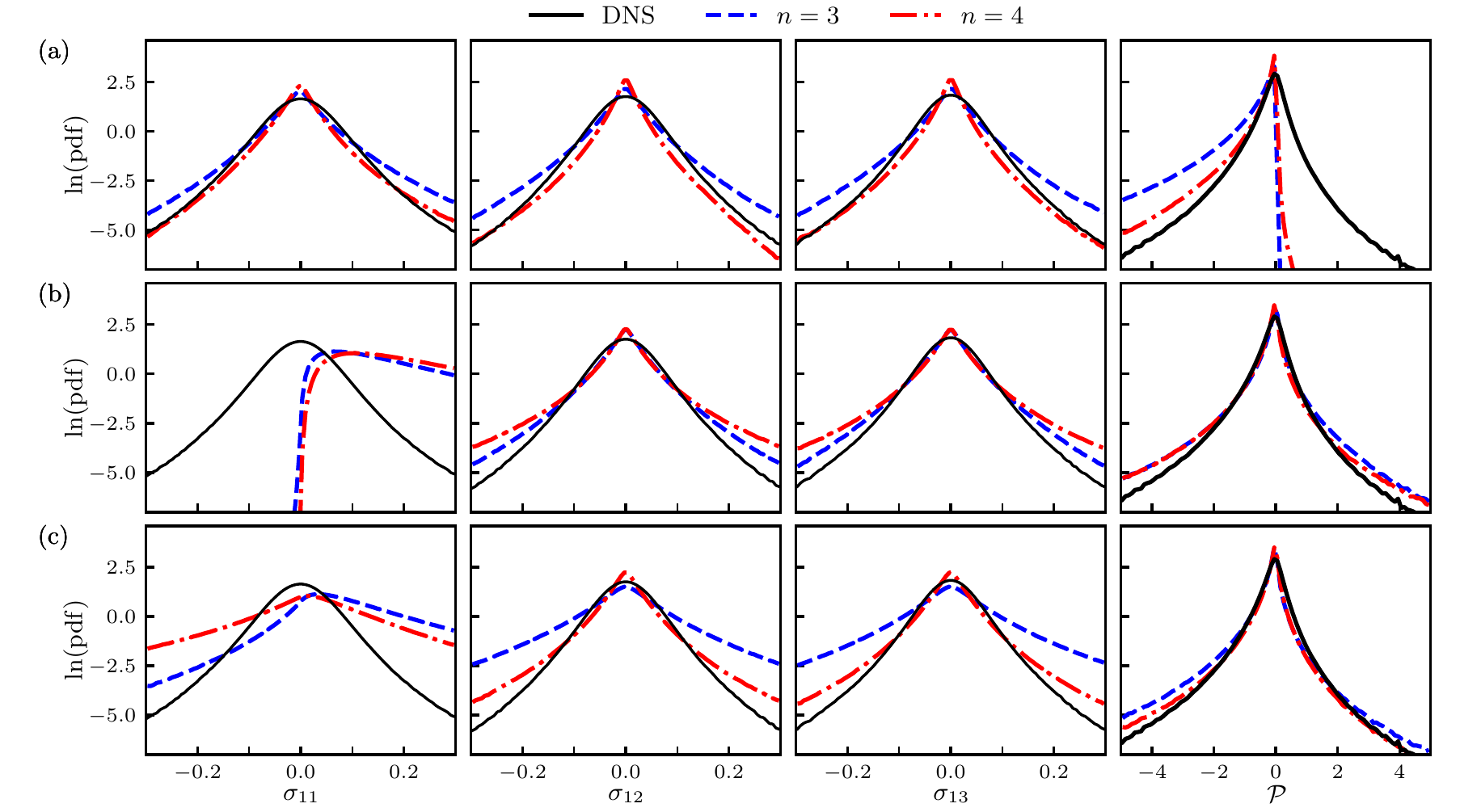}
	\caption{Modeled and reference summary statistics (i.e., pdfs) for $\sigma_{11}$ (first column), $\sigma_{12}$ (second column), $\sigma_{13}$ (third column), and $\mathcal{P}$ (last column). Reference results (solid black lines) are obtained from DNS of HIT \cite{li2008} and modeled results are obtained using the MAP parameter estimates given in Table \ref{tab:MAP} for (a) $P^{(n)}_{\sigma}(\bm{c})$, (b) $P^{(n)}_{\mathcal{P}}(\bm{c})$, and (c) $P^{(n)}_{\sigma\mathcal{P}}(\bm{c})$. Results are shown for both three ($n=3$; blue dashed lines) and four ($n=4$; red dash-dot lines) models given by Eq.~\eqref{eq:closure}.}
	\label{fig:compare_sum_stat}
\end{figure*}
% 	%figure-------------------------------

For the four-parameter model, Figure~\ref{fig:4_params} shows that $M^{(4)}_\mathcal{D}(c_1)$ is largely similar to the corresponding marginal distribution for the three-parameter model, with MAP values of roughly $c_1^\mathrm{MAP}=-0.04$ for each type of data (see Table \ref{tab:MAP}). All two-dimensional joint pdfs remain symmetric with respect to $c_2$, as in the three-parameter case, and there are large differences in $c_3^\mathrm{MAP}$ and $c_4^\mathrm{MAP}$ depending on whether stress or production data, or a combination of these data types, are used in the ABC-IMCMC approach. Once more, the combined results are generally most similar to the results for the production data alone, although there are large differences in $c_4^\mathrm{MAP}$ for the production only and combined data. 

Finally, in the four-parameter model, $c_3$ and $c_4$ are strongly correlated, as shown in Figure~\ref{fig:4_params}(l) for $P^{(4)}_{\sigma}(c_3,c_4\,|\, c_1^\mathrm{MAP},c_2^\mathrm{MAP})$. This strong correlation is a good example of when the proposal kernel with adaptive covariance (see Section~\ref{sec:method_adaptive}) can improve the acceptance rate of the ABC-MCMC algorithm.

It should be noted that the values of $c_1^\mathrm{MAP}$ summarized in Table \ref{tab:MAP} can be connected to the classical Smagorinsky coefficient as $C_\mathrm{S}=(-c_1^\mathrm{MAP}/2)^{1/2}$. The resulting values of $C_\mathrm{S}$ then range from $C_\mathrm{S}=0.19$ for the three-parameter model using stress data to $C_\mathrm{S}=0.13$ for the four-parameter model using combined stress and production data. These values are in generally close agreement with the classical range of $C_\mathrm{S}$ of between 0.1 and 0.2 \cite{meneveau2000}. 

%===============================================
% Subsec: A Priori Testing
%===============================================
\subsection{\emph{A priori} tests\label{sec:apriori}}

The primary outcome of any particular ABC-IMCMC test is an estimate of the posterior joint pdf, such as those shown in Figures~\ref{fig:3_params}~and~\ref{fig:4_params}. From these posterior distributions, estimates of the most probable parameter values, as well as their uncertainties, can be obtained. 

Here we use the maxima of the estimated posteriors (i.e., the MAP values, summarized in Table \ref{tab:MAP}), as the most probable parameter values, and compute the resulting SGS stress and production summary statistics (i.e., pdfs) from the model in Eq.~\eqref{eq:closure}. Figure~\ref{fig:compare_sum_stat} shows these modeled summary statistics, denoted $\mathcal{S}'_{ij}$ and $\mathcal{S}'_\mathcal{P}$, respectively, for both $P^{(3)}_{\mathcal{D}}(\bm{c})$ and $P^{(4)}_{\mathcal{D}}(\bm{c})$, along with the corresponding reference statistics $\mathcal{S}_{ij}$ and $\mathcal{S}_\mathcal{P}$. In general, across all types of reference data $\mathcal{D}$, the performance of the four-parameter model is improved compared to the three-parameter model, as indicated by the improved agreement between the modeled and reference summary statistics. This demonstrates that the fourth basis function, $\widetilde{G}^{(4)}_{ij}$, is important for improving the accuracy of the second-order SGS stress model given by Eq.~\eqref{eq:closure}. 

For ABC-IMCMC using SGS stress data, Figure~\ref{fig:compare_sum_stat}(a) shows that, despite good agreement between the modeled [using the MAP parameter estimates from $P^{(3)}_{\sigma}(\bm{c})$ and $P^{(4)}_{\sigma}(\bm{c})$] and reference stress pdfs, neither the three- or four-parameter models correctly reproduces the positive values of the production (i.e., energy backscatter) observed in the reference data. This limitation is caused by using only SGS stresses during ABC-IMCMC, without explicitly taking into account the correct modeling of the SGS production process. By contrast, Figure~\ref{fig:compare_sum_stat}(b) shows that, when using MAP parameters from $P^{(3)}_{\mathcal{P}}(\bm{c})$ and $P^{(4)}_{\mathcal{P}}(\bm{c})$, both models are able to match the SGS production refernce pdf reasonably well, but the stress pdfs are in much worse agreement, particularly for the diagonal elements of the tensor. This thus shows that using only SGS production reference data during ABC-IMCMC is not sufficient to correctly reproduce the stresses using either the three- or four-parameter models.

Finally, an intermediate model where the MAP parameter estimates are taken from $P^{(3)}_{\sigma\mathcal{P}}(\bm{c})$ and $P^{(4)}_{\sigma\mathcal{P}}(\bm{c})$ is shown in Figure~\ref{fig:compare_sum_stat}(c). In this case, the model and reference results are in reasonably close agreement for both the SGS stress and production pdfs, with generally better agreement for the production pdfs. Other combined models could be obtained from ABC-IMCMC by using different weightings of the stress and production components of the combined distance function in Eq.~\eqref{eq:dist_combined}, but the present weighting does provide reasonable agreement between the modeled and reference statistics for the stresses and production simultaneously. 

It should be noted that the present results generally indicate that both three- and four-term second order models have difficulties in simultaneously capturing stresses and production. This difficulty has been observed previously for other models \cite{meneveau2000}, and ABC-IMCMC effectively provides a user-definable blend of accuracy for these given models. Obtaining even better simultaneous agreement for both the stress and production statistics requires a more sophisticated model, most likely with a greater number of degrees of freedom.

%===============================================
% Subsec: A Posteriori Testing
%===============================================
\subsection{\emph{A Posteriori} tests\label{sec:les_aposteriori}}
Ultimately, the SGS model parameter values estimated using ABC-IMCMC are intended for use in forward LES runs of both idealized and practically relevant flows. Here we show that the present MAP parameter estimates (summarized in Table \ref{tab:MAP}) result in two new stand-alone SGS models based on Eq.~\eqref{eq:closure} that permit stable solutions for HIT in forward LES runs.

These \textit{a posteriori} tests were performed using \verb|spectralLES|, a pseudospectral LES solver for model testing and development written in pure Python. The solver is based on the open-source, pure Python code, \verb|spectralDNS| \cite{mortensen2016}. The LES were initialized using a random initial velocity field with a prescribed isotropic energy spectrum, and turbulence was sustained using spectrally-truncated linear forcing of wavenumber shells $k=2$ and $k=3$. The resulting simulation thus produced statistically stationary HIT. Simulations were performed using the same domain, energy injection rate, viscosity, and LES filter scale as the DNS reference data used in the \textit{a priori} tests described in the previous section, with a $64^3$ uniform grid. For comparison, we also ran a simulation using the static Smagorinsky model with the standard Smagorinsky coefficient $C_\mathrm{S} = 0.1$.

\begin{figure}[t!]
	\centering\includegraphics[]{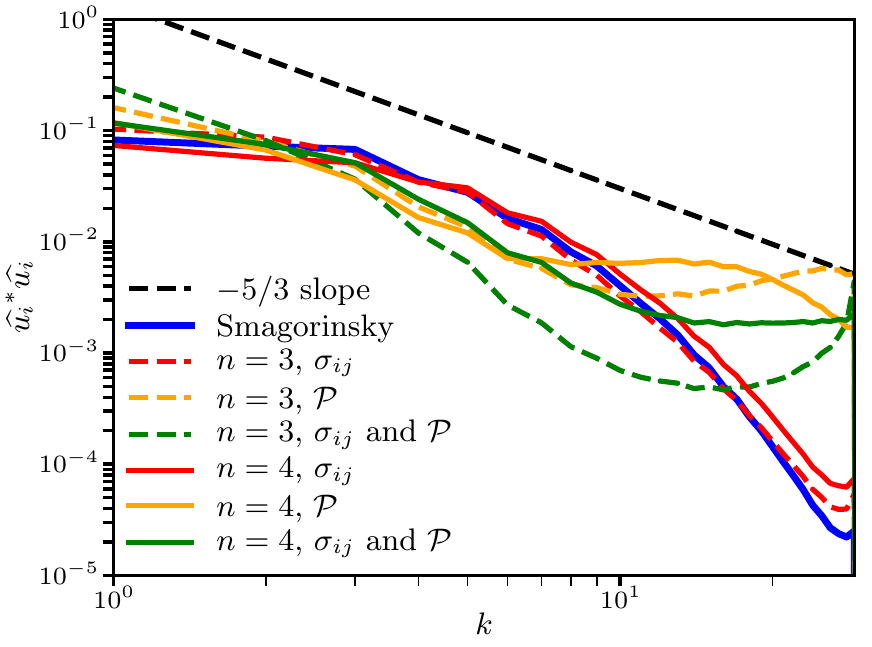}
	\caption{Kinetic energy spectra resulting from forward LES of HIT using the MAP values in Table \ref{tab:MAP} for the three- and four-parameters SGS models from Eq.~\eqref{eq:closure}. Each line is labeled by the number of parameters in the model (either $n=3$ or 4), and by the data used during the ABC-IMCMC procedure (either $\sigma_{ij}$, $\mathcal{P}$, or their combination). An LES spectrum obtained using the Smagorinsky model (solid blue line) and a $k^{-5/3}$ slope line are also shown.}
	\label{fig:spectra_best}
\end{figure}

Figure~\ref{fig:spectra_best} shows the resulting kinetic energy spectra from the LES using both three- and four-parameter models with parameters obtained from the MAP estimates in Table~\ref{tab:MAP}. Parameters determined using stress data result in spectra similar to the spectrum obtained using the Smagorinsky model. By contrast, spectra from LES with parameters based on the SGS production data tend to have larger magnitudes at higher wavenumbers. This is caused by the fact that, when using production reference data during ABC-IMCMC, parameters are chosen such that the modeled and reference production match at the LES scale.

%figure-------------------------------
%figure------------------------------- 

%figure-------------------------------
\begin{figure*}[t!]
	\centering\includegraphics[width=\linewidth]{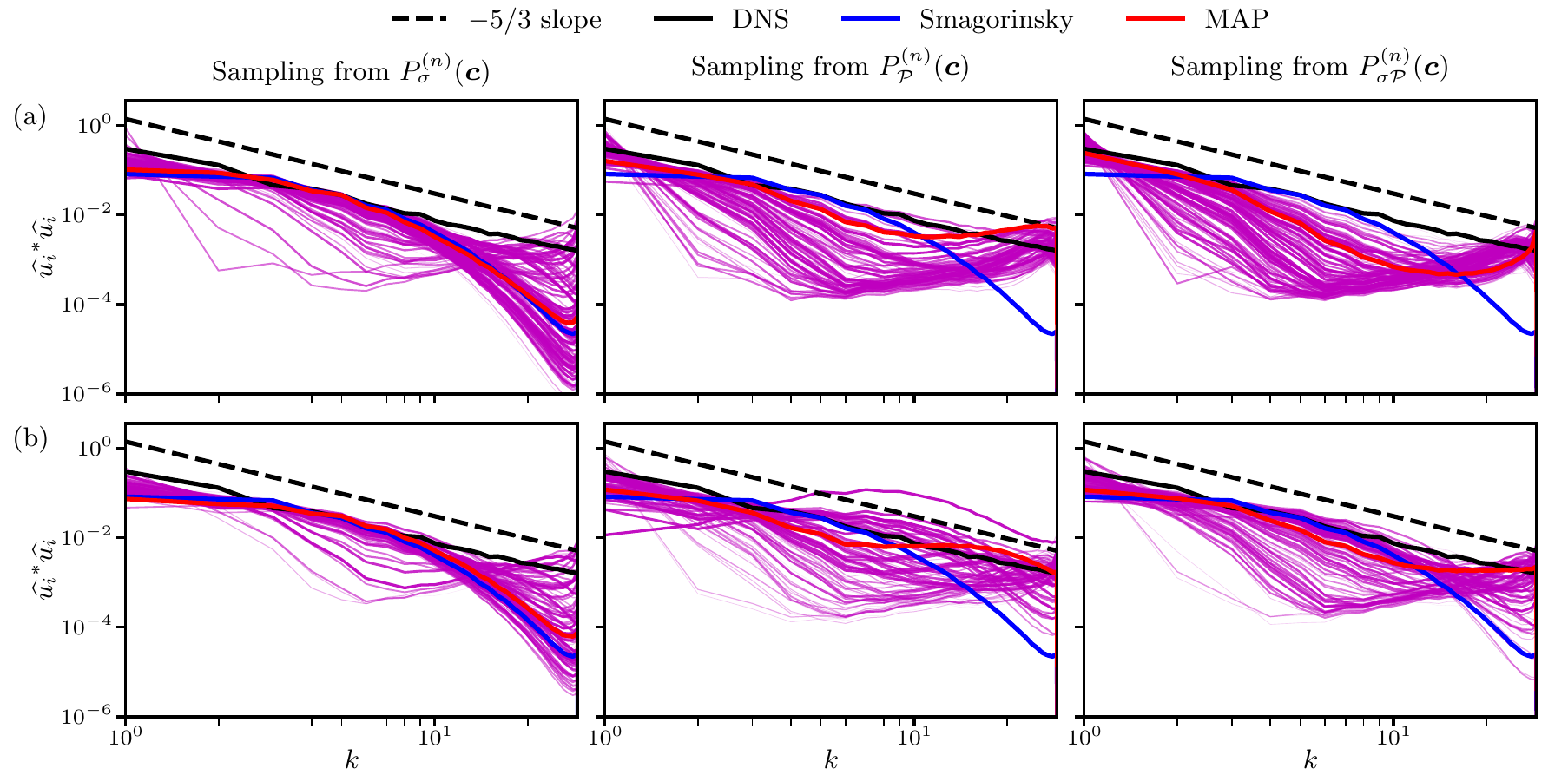}\vspace{-0.2in}
	\caption{Kinetic energy spectra (magenta lines) resulting from forward LES of HIT using parameter values sampled from $P^{(n)}_{\sigma}(\bm{c})$ (left column), $P^{(n)}_{\mathcal{P}}(\bm{c})$ (middle column), and $P^{(n)}_{\sigma\mathcal{P}}(\bm{c})$ (right column) for the (a) three- and (b) four-parameter SGS models from Eq.~\eqref{eq:closure}. The intensity of the magenta lines indicates the posterior probability of each sampled parameter set. Modeled spectra obtained using MAP parameter estimates (solid red lines), an LES spectrum obtained using the Smagorinsky model (solid blue lines), the reference data spectrum from DNS of HIT \cite{li2008} (solid black lines), and a $k^{-5/3}$ slope line are also shown.}
	\label{fig:forward_spectra}
\end{figure*}
% 	%figure------------------------------- 

The correspondence between the modeled and reference spectra when using production data during ABC-IMCMC parameter estimation is shown more clearly in Figure~\ref{fig:forward_spectra}. In this figure, we explore the propagation of uncertainty resulting from the various estimated posteriors by sampling 100 parameter sets from each of the six posterior pdfs using a Monte Carlo acceptance-rejection method. We then performed LES of HIT for each parameter set, resulting in the range of spectra shown in Figure~\ref{fig:forward_spectra} for each posterior. 

Figure~\ref{fig:forward_spectra} shows that there is substantial spread in the spectra for high wavenumbers when using parameters sampled from $P^{(3)}_{\sigma}(\bm{c})$ and $P^{(4)}_{\sigma}(\bm{c})$, and that the spectra are frequently overly dissipative at small scales. This yields better agreement of the modeled spectra with the purely dissipative Smagorinsky model than with the reference DNS results. By contrast, the spectra obtained by sampling from $P^{(3)}_{\mathcal{P}}(\bm{c})$ and $P^{(4)}_{\mathcal{P}}(\bm{c})$ are generally close to the DNS spectrum at small scales, with less variability at high wavenumbers. However, compared to spectra for the three-parameter model, the four-parameter model spectra are generally relatively close to the DNS spectrum over all wavenumbers. Finally, as with many of the other results presented herein, the spectra obtained by sampling the combined posteriors $P^{(3)}_{\sigma\mathcal{P}}(\bm{c})$ and $P^{(4)}_{\sigma\mathcal{P}}(\bm{c})$ are a blend of the spectra obtained from posteriors based on either the stresses or production alone.

%%%%%%%%%%%%%%%%%%%%%%%%%%%%%%%%
% Sec: Conclusions and Future Work
%%%%%%%%%%%%%%%%%%%%%%%%%%%%%%%%
\section{Conclusions}
In this study, we used ABC and IMCMC to estimate posterior distributions of unknown parameter values in three- and four-parameter SGS closure models for LES of turbulent flows. After outlining the general ABC-IMCMC approach, including sample algorithms, we used the method to estimate posterior distributions of model parameters based on pdfs (i.e., summary statistics) of SGS stress, SGS production, and a combination of SGS stresses and production. The reference data were obtained from DNS of HIT \cite{li2008}. Through \emph{a priori} tests, we showed that the estimated parameter values can be used to accurately reproduce stress and production pdfs, although simultaneously matching all pdfs was found to be difficult due to limitations of the model form. Using \emph{a posteriori} tests, we further showed that the models obtained using ABC-IMCMC can be stably integrated in forward LES runs using the \texttt{spectraLES} code.

It is emphasized that the present tests using nonlinear SGS models are a demonstration of the ABC-IMCMC approach for model parameter estimation, and that more complicated models, other types of reference data, and different choices of summary statistics and distance functions can be equally applied. The model evaluations that are part of the ABC-IMCMC approach can also be obtained from forward runs of an LES model, and the approach can be extended to the estimation of RANS model parameters, as described in \citet{doronina2019}. 

In order to ensure even greater accuracy when comparing modeled and reference results, more sophisticated models involving a greater number of unknown parameters are likely to yield improved simultaneous agreement for both stresses and the production. For example, \citet{king2016} used a very high-dimensional form of the SGS stress model to show that both the pdfs and the pointwise fields of the stresses and production can be reproduced. It is possible that lower dimensional models than those examined by \citet{king2016}, but with more than the four unknown parameters in the models examined here, may also retain some of the same accuracy. It is also possible that the greater computational efficiency enabled by ABC-IMCMC may allow the use of this approach with autonomic closure \cite{king2016}; an earlier effort focused on combining ABC with autonomic closure did not make use of the IMCMC method outlined here, and instead relied on the much less efficient ABC rejection sampling approach in Algorithm \ref{alg:abc-rej} \cite{doronina2018}.

Finally, we note that the calibration step, adaptive proposal, and MCMC procedure have all been included in order to accelerate the ABC process and reduce the requirement for computational resources during the model parameter estimation. However, further improvements to the ABC-IMCMC approach are also possible, including the use of linear regression to correct biases introduced due to the use of nonzero values of the tolerance $\varepsilon$. These and other techniques to further accelerate ABC-IMCMC and reduce the computational cost are important directions for future research. 

%%%%%%%%%%%%%%%%%%%%%%%%%%%%%%%%
% Sec: Acknowledgements
%%%%%%%%%%%%%%%%%%%%%%%%%%%%%%%%
\section*{Acknowledgements}
OAD and PEH acknowledge financial support from NASA award NNX15AU24A-03. PEH was also supported, in part, by AFOSR award FA9550-17-1-0144. Helpful discussions with Profs.\ Werner J.A.~Dahm, Ian Grooms, Will Kleiber, and Greg Rieker, as well as with Drs.~Jason Christopher and Scott Murman, are gratefully acknowledged.

%%%%%%%%%%%%%%%%%%%%%%%%%%%%%%%%
% Sec: References
%%%%%%%%%%%%%%%%%%%%%%%%%%%%%%%%
% \bibliographystyle{abbrvnat}
\bibliographystyle{unsrtnat}
\bibliography{reference_abc,biblio_abc} 

\end{document}